# Efficiency-Enhanced Open Earbud Earphone Antenna Using Dual-Feed Technique

SHIMING LIU[1], JIANHUA XIE[2], AND YAN WANG[1], (Senior Member, IEEE)

[1]Key Laboratory for Information Science of Electromagnetic Waves, College of Future Information Technology, Fudan University, Shanghai 200433, China.
[2] Shenzhen Shokz Co., Ltd., Shenzhen 518108, China.

CORRESPONDING AUTHOR: Y. WANG (e-mail: yanwang_fd@fudan.edu.cn).

This work was supported by the National Natural Science Foundation of China under Grant 62101133 and in part by Shanghai Rising Star Program under Grant 22QC1400100. (Corresponding author: Yan Wang).

**ABSTRACT** The stringent spatial constraints and the demand for high antenna efficiency in modern wireless earphones present significant design challenges. To address these issues, this paper presents and thoroughly investigates a novel earphone antenna design specifically tailored for open earbud wireless earphones. In contrast to traditional earphone antennas that rely on a conventional single-feed configuration, the proposed design introduces a dual-feed excitation technique incorporating a controlled phase difference between the two feeds. This innovative feeding strategy effectively enlarges the equivalent radiating aperture, thereby enhancing the overall radiation efficiency of the antenna system. Experimental and simulation results demonstrate that the dual-feed approach yields an efficiency improvement exceeding 1 dB when compared with standard single-feed designs. Furthermore, the fabricated prototype achieves a –6 dB impedance bandwidth that fully encompasses the 2.4 GHz ISM band, ensuring stable wireless communication performance. The measured total efficiencies reach –8.5 dB in free space and –9.5 dB under on-head conditions. These results confirm that the proposed antenna successfully achieves high efficiency and reliable performance within the extremely limited volume of an earbud device, demonstrating strong potential for integration into next-generation compact wireless earphones.

## I. INTRODUCTION

IN modern life, wireless earphones have become a popular product for portable and private audio experiences, offering greater freedom than wired ones [1]. The antenna performance is critical to wireless reliability in compact wearable devices such as earphones [2], [3], [4], [5], [6], [7], [8], [9]. However, for a better user experience, the space for antenna design in wireless earphones is usually small, resulting in low efficiency. Therefore, designing a high-efficiency wireless earphone antenna within a limited space is promising.

In the open literature, earphone antenna designs can be mainly divided into three categories based on the type of earphones: ear-bud earphone in Fig. 1(a) [10], [11], [12], ear-bar earphone in Fig. 1(b) [13], [14], [15] and open earbud earphone in Fig.1(c) [16]. As shown in Fig. 1(a), an earbud earphone consists of a single earbud, where the antenna is typically integrated. An on-head efficiency of –12.7 dB is achieved by a curved-slot antenna in ear-bud earphones [10]. An efficiency of –5.3 dB is enhanced by a cavity-backed configuration through the utilization of the cavity's shielding effect [11]. Similarly, the antenna size is reduced by 60% by integrating the cavity into the ear canal, while an efficiency of –7.5 dB is maintained [12]. For the ear-bar type earphone shown in Fig. 1(b), where the structure includes both the earbud and the ear bar, the antenna is typically integrated within the ear bar. A hook-shaped ring antenna with a half-wavelength loop above the ground plane is proposed in [13]. A parasitic metal structure is used to reduce head exposure and improve efficiency by 2 dB [14]. An analysis of the antenna modes of TWS earphones is conducted, where an inductor is employed to tune the ground resonance, resulting in a 5.2 dB efficiency improvement [15]. Fig. 1(c) shows the currently popular open earbud wireless earphone which uses an open structure to reduce the risk of hearing damage and ear canal infections [16]. For the open earbud earphone, the antenna is usually designed inside the earbud where the antenna space is very limited [17].

From Table I, it can be observed that significant progress has been made in the design of antennas for earbud and ear-bar earphones, achieving high performance for practical applications. Herein, a high-efficiency open earbud earphone antenna within the limited space is achieved. Also, the dual-feed technique is applied to an open earbud earphone antenna to achieve an efficiency improvement of over 1 dB.

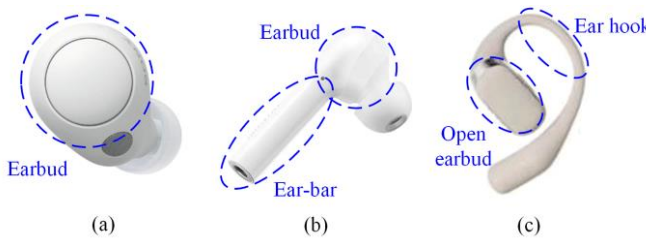

FIGURE 1. Main structure of three popular types of wireless earphones on the market. (a) Ear-bud earphone. (b) Ear-bar earphone. (c) Open earbud earphone.

TABLE 1. State-of-art of some typical earphone antennas

| Ref. | Type | Size[1] (mm$^3$) | Feed configuration | Efficiencies (dB) |
|---|---|---|---|---|
| [10] | Earbud | 20×20×1.5 | Single feed | -12.8 |
| [11] | Earbud | 33×33×10 | Single feed | -5.3(Peak) |
| [12] | Earbud | 26.1×26.1×14.7 | Single feed | -6.5(Peak) |
| [13] | Ear-bar | 35×17×4 | Single feed | N.A. |
| [14] | Ear-bar | 30×6.2×10 | Single feed | -5.1 |
| [15] | Ear-bar | 17.5×5×7.5 | Single feed | -6 |
| This work | Open earbud | 25×15×3.6 | dual feed | -9.5 |

[1] The size of the earphone antennas is represented as "length × width × height".

Previously, dual-feed methods have been used to mitigate local efficiency minima in inverted-F antennas (IFA) with parasitic branches by achieving in-phase current excitation [18] and to widen bandwidth in dual-element antennas with a 90° phase difference [19]. Zhang and Wang further explored differentially-driven microstrip antennas using an improved cavity model [20]. In this work, the dual-feed technique is introduced to enlarge the effective radiation area of a single patch, resulting in an over 1 dB improvement in on-head efficiency. Section II details the antenna configuration, Section III analyzes the operating mechanism, Section IV presents the measurements, and Section V concludes the paper.

## II. ANTENNA STRUCTURE AND DIMENSION

Fig. 2 presents the geometry and detailed dimensions of the proposed open earbud earphone antenna, which is composed of three primary parts: the patch antenna, the ground plane, and the feeding network. Substrate 1 (Sub 1) hosts the patch on its top layer and the ground plane on its bottom layer. The ground plane is closely attached to the top surface of Substrate 2 (Sub 2), whose bottom layer hosts the feeding network. As shown in Fig. 2(b), the patch antenna is positioned with its left region reserved for other components. The proposed design employs dual-feed excitation, implemented through feeding probes 1 and 2, which connect the feeding network to the patch for optimal performance. A $C_m = 1.1$ pF capacitor is loaded at the corner of the patch antenna to optimize impedance matching. Fig. 2(c) shows the feeding network, which consists of a phase shifter, a 3 dB power divider, and a matching circuit.

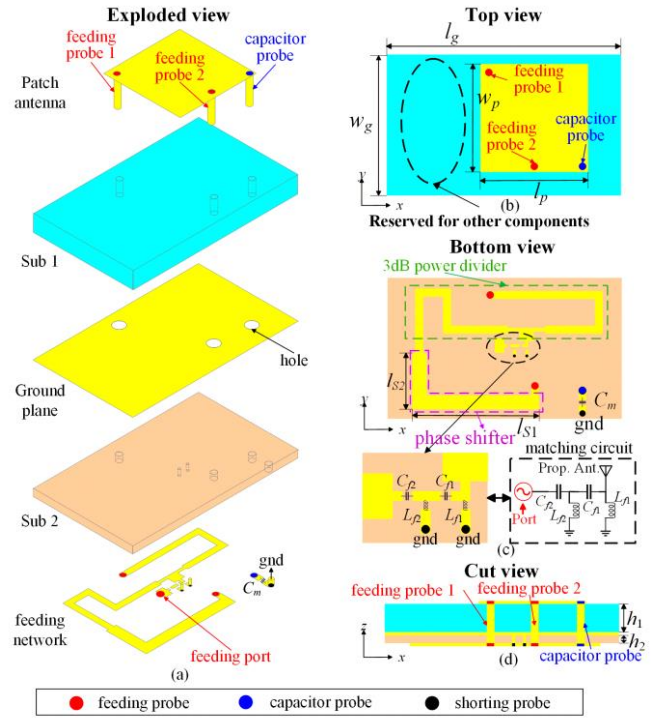

FIGURE 2. Geometry of the proposed antenna and its placement near the human head. (a) Exploded view. (b) Top view. (c) Bottom view. (d) Cut view. (e) The placement of the proposed antenna on the head and the position where the actual earphones are worn on the head. ($l_g$ = 25, $w_g$ = 15, $l_p$ = $w_p$ = 11.5, $l_{S1}$ = 12, $l_{S2}$ = 6, $h_1$ = 3, $h_2$ = 0.6, $d$ = 5.5, unit: mm)

The phase shifter is implemented as a 50 Ω microstrip line with a length of 18 mm and a width of 1.5 mm, which yields an 82° phase shift at 2.45 GHz. The 3 dB power divider includes two quarter-wavelength 70.7 Ω microstrip lines (0.85 mm wide), a 2 mm long 100 Ω line (0.4 mm wide), and a 2 mm long 50 Ω line. The matching circuit comprises two capacitors ($C_{f1}$ = 0.4 pF, $C_{f2}$ = 0.2 pF) and two inductors ($L_{f1}$ = 0.3 nH, $L_{f2}$ = 5.1 nH).

Sub1 is an FR-4 substrate with a relative permittivity of 4.4, a loss tangent of 0.025, and a thickness of 3 mm. In contrast, Sub2 is made of F4B material, characterized by a relative permittivity of 3, a loss tangent of 0.0015, and a thickness of 0.6 mm. Fig. 2(e) shows that the proposed antenna is designed to fit within the earbud of open earbud earphones. It is positioned on the head with the earbud oriented parallel to the head surface at a distance of $d$ = 5.5 mm, corresponding to the typical wearing position of actual wireless earphones.

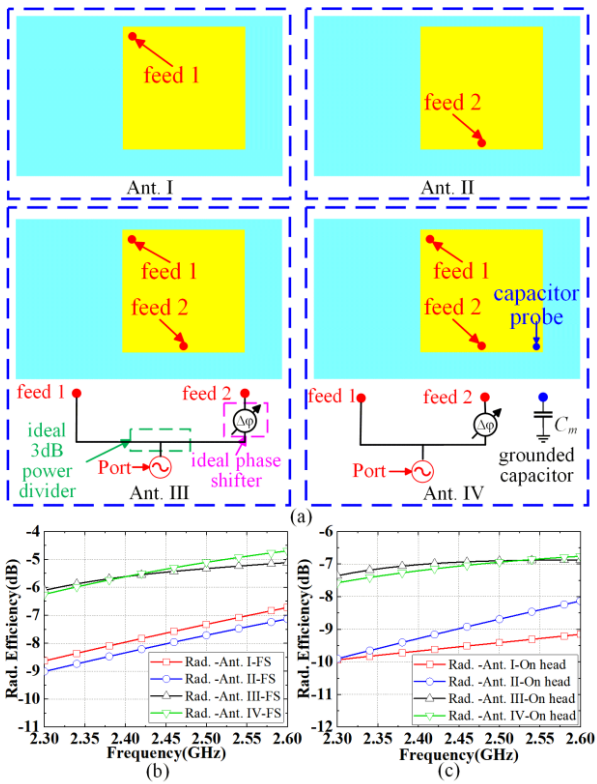

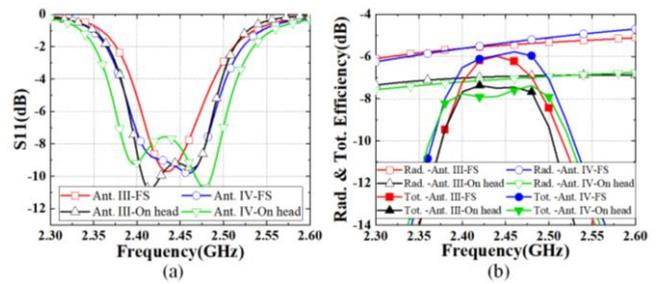

FIGURE 4. Simulated results of Ant. III, IV. (a) Simulated S11 of Ant. III, IV. (b) Simulated efficiencies of Ant. III, IV.

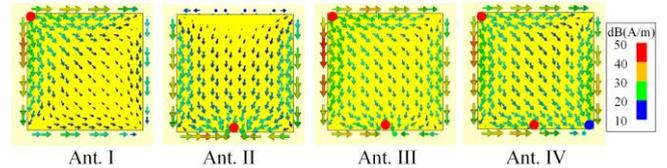

FIGURE 5. Simulated vector current distributions on the patches of Ant. I~ IV at 2.45GHz.

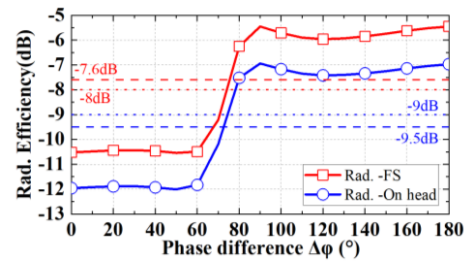

FIGURE 6. Simulated antenna radiation efficiencies with different feeding phase difference $\Delta\varphi$.

FIGURE 3. Antenna design process with radiation efficiency. (a) Antenna design process. (b) Simulated radiation efficiency of Ant. I~IV in free space. (c) Simulated radiation efficiency of Ant. I~IV on the head.

## III. ANTENNA DESIGN PROCESS

The study first outlines the antenna design and performance characteristics, followed by an investigation of the dual-feed mechanism and the impact of phase difference. Finally, the proposed dual-feed antenna design is presented.

### A. ANTENNA DESIGN PROCESS WITH PERFORMANCE

To illustrate the proposed dual-feed earphone antenna, the design process is presented in Fig. 3(a). Ant. I is a single-feed patch antenna excited at the corner (feed 1). Ant. II is modified from Ant. I by moving the feed to the midpoint of the lower edge (feed 2). Ant. III is formed by simultaneously exciting both feed 1 and feed 2 using an ideal 3 dB power divider and a phase shifter with phase difference $\Delta\varphi$. Ant. IV is derived from Ant. III by loading a capacitor at the corner.

Fig. 3(b) and (c) illustrate the simulated radiation efficiencies of Ant. I to Ant. IV in free space and on the head. Ant. I and Ant. II exhibit comparable efficiencies of –7.6 dB and –8.0 dB in free space, and –9.5 dB and –9.0 dB on the head, respectively. In contrast, Ant. III achieves significantly improved efficiencies of –5.5 dB in free space and –6.9 dB on the head, outperforming both Ant. I and Ant. II. These results demonstrate that dual-feed excitation effectively enhances radiation efficiency. In addition, Ant. IV exhibits similar radiation efficiency to Ant. III, indicating that the capacitor $C_m$ has a slight effect on the antenna radiation efficiency.

To illustrate the function of the capacitor $C_m$, Fig. 4 shows the performance of Ant. III and Ant. IV. Similar antenna designs that utilize capacitors for impedance matching optimization have been applied in published literature [15], [21], [22], [23], [24]. As shown in Fig. 4(a), Ant. IV exhibits a broader –6 dB impedance bandwidth than Ant. III. With the addition of a 1.4 pF capacitor ($C_m$), the on-head –6 dB bandwidth increases from 98 MHz (2390 MHz–2488 MHz) for Ant. III to 129 MHz (2375 MHz–2504 MHz) for Ant. IV. As illustrated in Fig. 4(b), the average radiation efficiencies of Ant. IV improved to –6.3 dB in free space and –7.7 dB on the head, compared to –6.9 dB and –8.0 dB for Ant. III.

### B. WORKING MECHANISM

Fig. 5 shows the simulated vector current distributions at 2.45 GHz on the patches of Ant. I to IV, illustrating the working mechanism of the dual-feed technique. As shown in Fig. 5, Ant. I and Ant. II operate in a quarter-wavelength mode, where the current is strong near the feeding point but weak farther away, resulting in a limited effective radiating area. By employing the dual-feed technique with a 90° phase difference, the excitation currents from the two ports can constructively combine, resulting in a notable enhancement. A more uniform and stronger current distribution typically corresponds to a larger effective radiating area and thus higher radiation efficiency. As a result, a high radiation efficiency is achieved.

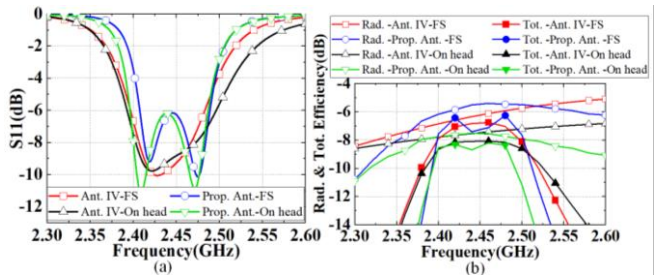

FIGURE 7. Simulated results of Ant. IV and Prop. Ant. (a) Simulated S11 of Ant. IV and Prop. Ant. (b) Simulated efficiencies of Ant. IV and Prop. Ant.

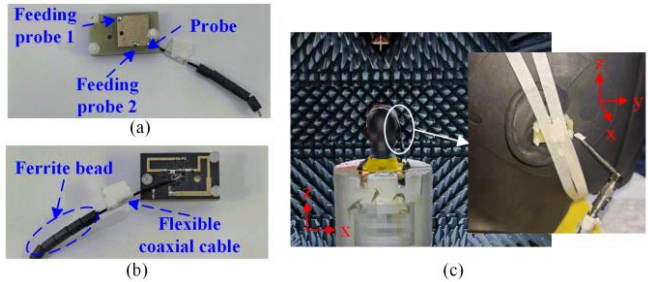

FIGURE 8. Photos of the antenna prototype and measurement setup in the anechoic chamber. (a) Top view. (b) Bottom view. (c) Anechoic chamber measurement environment and on-head placement of the antenna prototype during testing

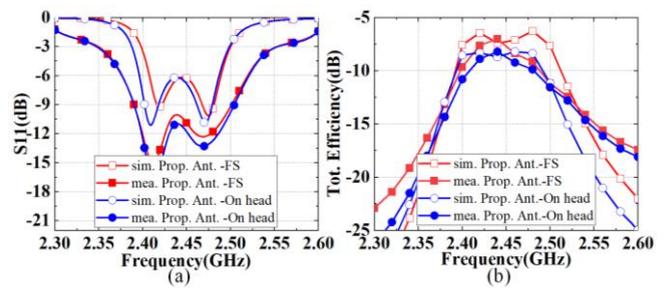

FIGURE 9. Comparison of simulated and measured performance of the Prop. Ant. in free space and on the head. (a) Reflection coefficient (S11). (b) Total efficiency.

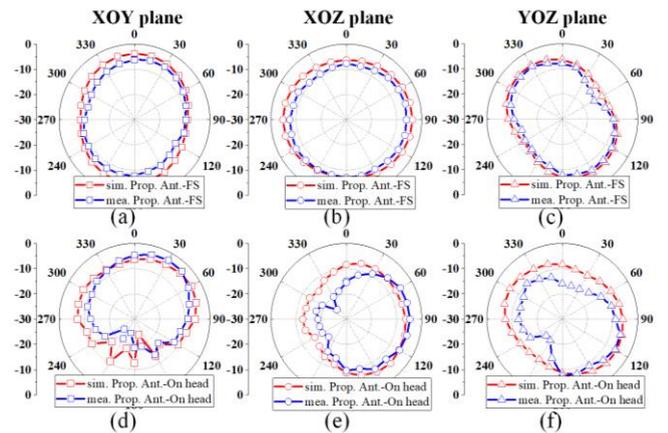

FIGURE 10. Simulated and measured radiation patterns of the Prop. Ant. at 2.45 GHz in free space and on the head. (a) XOY plane in free space. (b) XOY plane on the head. (c) XOZ plane in free space. (d) XOZ plane on the head. (e) YOZ plane in free space. (f) YOZ plane on the head.

To evaluate the impact of the phase difference ($\Delta\varphi$) between the two feeds on antenna efficiency, Fig. 6 presents the radiation efficiencies of Ant. III under various $\Delta\varphi$ values. When $0° \leq \Delta\varphi \leq 60°$, the radiation efficiency ranges from $-10.5$ dB to $-12$ dB in both free space and on the head. When $80° \leq \Delta\varphi \leq 180°$, the efficiency increases to approximately $-6.5$ dB, highlighting the importance of phase control in the dual-feed technique. Although the antenna exhibits performance similar to differential feeding, there are differences in excitation phase difference, main purpose, common-mode suppression level, and structural design. Overall, the proposed antenna aims to achieve current superposition and an increased effective radiating area through a dual-feed technique with a specific phase difference, resulting in over 1 dB improvement in radiation efficiency compared to single-feed antennas.

### C. DESIGN OF THE PROPOSED ANTENNA

To implement Ant. IV in Fig. 3, the phase shifter and power divider are realized using microstrip lines, as shown in Fig. 2. An 18 mm-long, 50 Ω microstrip line is used to generate the required phase difference between the two feeds, introducing a phase delay of approximately 82° at 2.45 GHz. The 3 dB power divider is constructed using microstrip lines with different characteristic impedances. Structural asymmetry and size constraints in the feeding network can cause power differences between the two ports, but its effect on antenna efficiency is negligible. The matching circuit consists of: a paralleled inductor $L_{f1}$, a serialized capacitor $C_{f1}$, a paralleled inductor $L_{f2}$, and a serialized capacitor $C_{f2}$. Fig. 7 compares the performance of Ant. IV and the proposed antenna. Over the 2400MHz–2500 MHz band, their on-head radiation efficiencies are $-7.4$ dB and $-7.5$ dB, respectively, with corresponding average system efficiencies of $-8.3$ dB and $-8.9$ dB. Due to the losses from the dielectric and the feeding network, the proposed antenna exhibits gain of $-5.1$ dBi in free space and $-3.9$ dBi on the head, respectively. The simulated SAR values of the proposed antenna under FCC and CE standards are 0.28 W/kg (1 g) and 0.09 W/kg (10 g), both below the regulatory limits. Therefore, a high-efficiency open earbud earphone antenna using dual-feed technique is achieved.

### IV. PROTOTYPE AND MEASUREMENT

A prototype is fabricated and fixed using nylon pillars. A flexible coaxial cable with six ferrite beads connects the antenna to an external source, with its outer and inner conductors soldered to the ground and feeding port, respectively. Fig. 8 shows the prototype and measurement setup in an anechoic chamber. The head model is a CTIA-compliant homogeneous silicone–carbon dielectric based on the SPEAG PG10 specific anthropomorphic mannequin (SAM) [25]. During testing, $C_m$ is 1 pF, and the matching circuit is tuned to $L_{f1} = 0.7$ nH, $C_{f1} = 0.5$ pF, $L_{f2} = 4.2$ nH and $C_{f2} = 0.3$ pF.

Fig. 9(a) presents the simulated and measured S-parameters of the proposed antenna (Prop. Ant.) in free space and on the head. The measured –6 dB impedance bandwidths are 134 MHz (2373 MHz–2507 MHz) in free space and 140 MHz (2380 MHz–2520 MHz) on the head, compared to simulated values of 82 MHz (2407 MHz–2489 MHz) and 91 MHz (2396 MHz–2497 MHz), respectively. Fig. 9(b) shows the corresponding efficiencies. In free space, the measured average efficiency over 2400 MHz–2500 MHz is –8.5 dB, while the simulated value is –7.1 dB. On the head, the measured and simulated average efficiencies are –9.5 dB and –8.9 dB, respectively. The discrepancy between free-space and on-head results is mainly attributed to boundary condition changes introduced by the human head [26], [27].

Fig. 10 presents the measured and simulated radiation patterns of the proposed antenna in free space and on the head, showing good agreement under both conditions. The measured gains are –5 dBi in free space and –2.4 dBi on the head, aligning well with simulation results. Discrepancies between simulated and measured results might be attributed to factors such as material property inaccuracies, feed-line coupling, and placement offsets of the antenna on the head.

## V. CONCLUSION

This paper investigates the application of a dual-feed technique with phase difference in earphone antennas to improve radiation efficiency. By enhancing the radiating current distribution and effective radiating area, the dual-feed technique achieves a notable improvement in radiation efficiency over conventional single-feed designs. The feeding network of the dual-feed antenna is realized using microstrip lines of different impedances. The prototype of the dual-feed antenna exhibits a measured -6 dB impedance bandwidth of 134 MHz (2373 MHz–2507 MHz) in free space and 140 MHz (2380 MHz–2520 MHz) on the head, with corresponding efficiencies of −8.5 dB and −9.5 dB, respectively. In summary, this paper presents a compact dual-feed wireless earphone antenna with enhanced radiation efficiency, showing strong potential for practical implementation.